\documentclass[aps,prb,twocolumn,groupedaddress,showpacs]{revtex4}%
\usepackage{amsmath}
\usepackage{graphics}
\usepackage{graphicx}
\usepackage{epsfig}
\usepackage{amssymb}
\usepackage{amsmath}%
\usepackage{amsfonts}

\begin{document}
\title{Bose-Einstein condensation in antiferromagnets close to the saturation field}
\author{D. \surname{Reyes}}
\affiliation{Centro Brasileiro de Pesquisas F\'{\i}sicas - Rua Dr.
Xavier Sigaud, 150-Urca, 22290-180,RJ-Brazil}
\author{A. \surname{Paduan-Filho} }
\affiliation{Instituto de F\'{\i}sica, Universidade de S\~ao Paulo,
S\~ao Paulo, Brazil}
\author{M. A. \surname{Continentino}}
\affiliation{Instituto de F\'{\i}sica, Universidade Federal Fluminense, \\
Campus da Praia Vermelha, Niter\'oi, RJ, 24.210-340, Brazil}

\date{\today}

\begin{abstract}
At zero temperature and strong applied magnetic fields the ground
sate of an anisotropic antiferromagnet is a saturated paramagnet
with fully aligned spins. We study the quantum phase transition as
the field is reduced below an upper critical $H_{c2}$ and the
system enters a XY-antiferromagnetic phase. Using a bond operator
representation we consider a model spin-1 Heisenberg
antiferromagnetic with single-ion anisotropy in hyper-cubic
lattices under strong magnetic fields. We show that the transition
at $H_{c2}$ can be interpreted as a Bose-Einstein condensation
(BEC) of magnons. The theoretical results are used to analyze our
magnetization versus field data in the organic compound
$NiCl_2$-$4SC(NH_2)_2$ (DTN) at very low temperatures. This is the
ideal BEC system to study this transition since $H_{c2}$ is
sufficiently low to be reached with static magnetic fields (as
opposed to pulsed fields). The scaling of the magnetization as a
function of field and temperature close to $H_{c2}$ shows
excellent agreement with the theoretical predictions. It allows to
obtain the quantum critical exponents and confirm the BEC nature
of the transition at $H_{c2}$.
\end{abstract}

\maketitle

The organic compound $NiCl_2$-$4SC(NH_2)_2$ (DTN) undergoes a field
induced non-magnetic to XY-antiferromagnetic transition
\cite{Paduan1,Paduan2}. This transition can be viewed as a
Bose-Einstein condensation of magnons associated with the Ni spin
$1$ degrees of freedom. Other magnetic systems with a singlet ground
state either with spin-1 Ni atoms or spin-1/2 dimers have also been
shown to exhibit this transition\cite{Batyev,Wessel,Nohadani}. At
zero temperature it is driven by the magnetic field $H$ that reduces
the Zeeman energy of the $S^z=1$ state until it becomes degenerate
with that of the product state of $S^z_i=0$. At this point,
$H=H_{c1}$, the antiferromagnetic (AF) interactions give rise to a
long range ordered phase. Experimentally, the magnetization $M$ at
very low temperatures starts to increase above the critical magnetic
field $H_{c1}$ and eventually saturates above a critical magnetic
field $H_{c2}$ \cite{Paduan1,Paduan2,Nikuni,Radu,Zapf,Sebastian}.
The transition at $H_{c1}$ has been intensively investigated, both
theoretically \cite{Wang2,Wang3} and
experimentally\cite{Nikuni,Radu,Paduan1,Paduan2,Zapf,Sebastian},
while the one at $H_{c2}$ is much less studied. The DTN  is the
ideal BEC system to investigate the latter transition since detailed
magnetization curves can be obtained close to the critical field
$H_{c2}=12.3$T. In other well known BEC systems, as
$BaCuSi_{2}O_{6}$\cite{Jaime} and $TlCuCl_3$\cite{rice}, the
critical fields $H_{c2}$ are $49$T and $83$T, respectively, and
presently can only be reached using pulsed fields. The excellent
quality of the magnetization versus field curves obtained in DTN
using standard superconducting coils is essential for the scaling
analysis presented here.

In this paper we study the transition at $H_{c2}$. A theoretical
approach is more directly developed, starting from the saturated
paramagnetic (PARA) phase. We consider decreasing the external
magnetic field at $T=0$ to the critical value $H_{c2}$ where the
transverse components of the magnetization condense. A scaling
approach for this transition has recently been
proposed\cite{Mucio4}. In this Communication we provide the
microscopic theory for this transition. We identify its universality
class as a Bose-Einstein condensation associated with a dynamic
exponent $z=2$. We compare the predictions for the scaling behavior
of the magnetization close to the quantum critical point (QCP)
($T=0$, $H=H_{c2}$) with experimental magnetization data on DTN and
obtain an excellent agreement.

For a long time the magnetically ordered state and low energy
excitations of quantum Heisenberg magnets have been studied using
the spin-wave expansion (see \cite{Keffer} and references
therein). This is usually implemented by expressing the components
of the spin operators at lattice sites $i$ in terms of canonical
boson operators $b_{k}^{\dagger}$ and $b_k$ using the
Holstein-Primakoff (H-P) transformation\cite{HP},  the
Dyson-Maleev (D-M) transformation\cite{Dyson,Maleev}, or the
Schwinger transformation\cite{Schwinger} (ST). Here we use the
bond-operator mean-field theory (BOMFT) \cite{sachdev} to study
spin-1 Heisenberg AF in hyper-cubic lattices with single-ion
anisotropy close to the quantum phase transition at the saturation
field $H_{c2}$. It yields the phase diagram and the thermodynamic
behavior of the model close to $H_{c2}$. The BOMFT gives an exact
description of this transition for three dimensional (3d) systems.
The reason is that the effective dimension $d_{eff}=d+z=5$
associated with the QCP is larger than the upper critical
dimension $d_c=4$ above which mean-field theory is exact.

The Hamiltonian describing the magnetic system is,
\begin{equation}\label{hamilt}
\mathcal{H}\!=\!\frac{J}{2}\!\sum_{\langle
i,j\rangle}(S_{i}^{x}S_{j}^{x}\!+\!
S_{i}^{y}S_{j}^{y}\!+\!S_{i}^{z}S_{j}^{z})+D\!\sum_{i}(S_{i}^{z})^{2}
\!-\!H\!\sum_{i}S_{i}^{z}
\end{equation}
where the sum is over all nearest neighbor pairs of a
$d$-dimensional hyper-cubic lattice with $N$ sites occupied by spins
with $S=1$. $J>0$ is the AF exchange coupling, $D$ is the single-ion
anisotropy, $H$ the magnetic field applied in the $z$-direction ($g
\mu_B=1$). Starting from the bond-operator representation for two
spins $S=1/2$\cite{sachdev}, Wang and
collaborators\cite{Wang2,Wang3,Wang1} obtained a representation for
a spin-1 Heisenberg system with a single-ion anisotropy in terms of
these operators. At zero temperature and for external magnetic
fields larger than the saturation magnetic field $H_{c2}$ the spins
are fully aligned with the field. In this case the bond operator
representation can be expressed as,
\begin{equation}\label{HP}
S^{+}=\sqrt{2}\bar{u}t_z,
\hspace{0.5cm}S^{-}=\sqrt{2}\bar{u}t_{z}^{\dagger},
\hspace{0.5cm}S^{z}=1-t_{z}^{\dagger}t_{z}
\end{equation}
with the constraint
$u_{i}^{\dagger}u_{i}+t_{i,z}^{\dagger}t_{i,z}=1$. We have used that
in this large field case the components of the spins perpendicular
to the field can be projected out and those parallel {\it condense},
such that, $u_i=u_{i}^{\dagger}=\bar{u}$. This mapping is exact for
$H>H_{c2}$, where the probability of the down spin state
$d^{\dagger}d|0\rangle$ is strictly zero at $T=0$. The magnetic
ordering for $H=H_{c2}$ can be identified as a Bose-Einstein
condensation of the transverse components of the spins which give
rise to the collective \textbf{magnon} excitations.  The operators
$t_{i,z}^{\dagger}t_{i,z}$ describe the departure of the spins from
the field direction and are associated with these excitations.
Replacing Eq. ({\ref{HP}}) in Eq. ({\ref{hamilt}}) with
$S^{\pm}=S^{x}\pm iS^{y}$ and changing from atomic to normal
coordinates, i.e., taking $t_{i,z}^{\dagger}=
 \frac{1}{\sqrt{N}} \sum_{k}e^{-ik.r_{i}}  b_{k}^{\dagger}$, we
 have,
\begin{equation}\label{mf}
\mathcal{H}_{mf}=\sum_{k}\omega_{k} b_{k}^{\dagger}b_{k}+E_{g},
\end{equation}
where,
\begin{equation}\label{eg}
E_{g}=N\left(\frac{JZ}{2}+D\bar{u}^{2}-H-\mu\bar{u}^{2}+\mu\right)
\end{equation}
is the ground state energy of the system. The dispersion relation of
the excitations for $H>H_{c2}$ is given by,
\begin{equation}\label{w}
\omega_k=H-(D+\mu)-JZ(1-\bar{u}^{2}\gamma_k)
\end{equation}
with,
$\gamma_k=d^{-1}\sum_{\nu=1}^{d}\cos(\mathbf{k}.\mathbf{a}_{\nu})$
and $Z$ the number of nearest neighbors. The chemical potential
$\mu$ was introduced to impose the constraint condition of single
occupancy. This and the parameter $\bar{u}$ are determined by
solving coupled, self-consistent, saddle point
equations\cite{Wang3}.

For fixed field $H$  or temperature $T$, the thermodynamic
quantities can be obtained from the internal energy $U$. This is
given by\cite{D1,D2}:
$U=\sum_{k}\omega_{k}<b_{k}^{\dagger}b_{k}>+E_{g}$, with the Bose
factor, $n_k=<b_{k}^{\dagger}b_{k}>=
\frac{1}{2}\left(\coth\frac{\beta \omega_k}{2}-1\right)$ and $E_g$
given by Eq. (\ref{eg}). For simplicity, we impose boundary
conditions on a $d$-dimensional hyper-cubic lattice with primitive
lattice vectors $\mathbf{a}_{\nu}$ and lattice spacing
$a=|\mathbf{a}_{\nu}|=1$. The minus sign in front of $J$ takes into
account explicitly the antiferromagnetic nature of the exchange
interactions.  For hyper-cubic lattices, the minimum of the
spin-wave spectrum occurs for $q=Q=(\pi/a,\pi/a,\pi/a)$ in three
dimensions. Since $\omega_{Q}=H-(D+\mu)-JZ(\bar{u}^{2}+1)$, the
condition $\omega_{Q}\equiv0$ defines the critical field
$H_{c2}=D+\mu+JZ(\bar{u}^{2}+1)$, below which the spin-wave energy
becomes negative signaling the entrance of the system in the AF
phase. Finally, writing $k=Q +q$ and expanding for small $q$, the
spin-wave dispersion relation can be obtained as follows:
\begin{equation}\label{wh}
\omega_{q}=(H-H_{c2})+\mathfrak{D}q^{2},
\end{equation}
where the spin-wave stiffness at $T=0$ is given by,
$\mathfrak{D}=J\bar{u}^{2}$. The quantum phase transition at
$H_{c2}$ has a dynamic exponent $z=2$ due to the
ferromagnetic-like dispersion of the magnons, in spite of the
antiferromagnetic character of the interactions\cite{Mucio4}.

\emph{Magnetization}---Close to the critical field $H_{c2}$ the
temperature-dependent magnetization should follow  a power
law\cite{Mucio3}. We define the variation of uniform magnetization
per volume $V$ as, $\Delta
M=(M_{sat}-M)/V=\sum_{k}<b_{k}^{\dagger}b_k>$ where $M_{sat}$ is the
saturation magnetization. Considering the spectrum of excitations,
Eq. (\ref{wh}), we have in the thermodynamic limit,
\begin{equation}\label{m}
\Delta M\!=\!\frac{S_{d}}{4\pi^{d}
\mathfrak{D}^{d/2}}(k_{B}T)^{d/2}\!\!\int_{y}^{\infty}\!\!\!\!dx\left(x\!-\!y\right)^{\frac{d}{2}-1}\!\left(\coth\frac{x}{2}\!-\!1\right)
\end{equation}
where $x=\beta \omega_q=y+\beta^{2}\mathfrak{D}q^{2}$, $y=\beta
\delta$, $S_{d}$ the solid angle and we have defined
$\delta=|H-H_{c2}|$ as the distance to the QCP. At this point, we
have to consider in which region of the phase diagram (see Fig.1 of
Ref.\cite{Mucio4}) we are interested. Because we want to calculate
the magnetization above $H_{c2}$ and particularly at the quantum
critical trajectory, $H=H_{c2}$, $T \rightarrow 0$, we consider
region II in Figure 1 of Ref.\cite{Mucio4} where $k_BT\gg \delta$
and consequently $y\ll1$. Calculating the integral above in three
dimensions, we obtain, $\Delta M_{3d}=$polylog$\left(\frac{3}{2},
e^{-y}\right)(k_{B}T)^{3/2}/\left(\pi^{3/2}\mathfrak{D}^{3/2}\right)$,
where polylog$(a,z)=\sum_{n=1}^{\infty} z^{n}/n^{a}$ is the general
polylogarithm function of index $a$ at the point $z$. Along the
quantum critical trajectory, $H=H_{c2}$, $T \rightarrow 0$, we find
$\Delta
M_{3d}(\delta=0)=g\mu_{B}\zeta(3/2)(k_{B}T)^{3/2}/\left(\pi^{3/2}\mathfrak{D}^{3/2}\right)$,
where $\zeta$ is the Riemann zeta-function.

\emph{Specific heat}---From the internal energy $U$ obtained before
and using the thermodynamic relation, $C_V=\partial U/\partial T$ we
get,
\begin{equation}
C_V=\frac{ k_{B}S_{d}   (k_{B}T)^{d/2}     }{2\pi^{d}
\mathfrak{D}^{d/2}} \int_{y}^{\infty}dx
x^{2}\left(x-y\right)^{\frac{d}{2}-1}\sinh^{-2}\left(\frac{x}{2}\right).
\end{equation}
Again, we consider $y\ll 1$ and along the quantum critical
trajectory in $3d$, we find, $C_{V}(\delta=0)=
15k_{B}(\sqrt{1+4\pi}-1)\left(k_{B}T\right)^{3/2}/(8\pi^{3/2}\mathfrak{D}^{3/2})$.

\emph{Susceptibility}--- Here we use the relation, $\chi=\partial M/
\partial H$, where $M$ is the magnetization. Taking
the derivative of Eq. (\ref{m}) with respect to $H$ and changing
variables, we have at $\delta=0$,
\begin{equation}\label{x}
\chi=\frac{S_{d}}{8\pi^{d}
\mathfrak{D}^{d/2}}(k_{B}T)^{\frac{d}{2}-1}
\int_{0}^{\infty}dxx^{\frac{d}{2}-1}\sinh^{-2}\left(\frac{x}{2}\right).
\end{equation}
For $3d$ the longitudinal susceptibility at the critical field is
given by,
$\chi(\delta=0)=(k_{B}T)^{1/2}/((\pi-1)\sqrt{\pi}\mathfrak{D}^{3/2})$.

The temperature dependence of the physical properties calculated
above in $3d$, can be  easily obtained from the fact that the QCP at
$H_{c2}$ is governed by \emph{Gausssian} exponents and the free
energy has a scaling form\cite{Mucio3},
\begin{equation}
\label{free}
 f \propto |\delta |^{2-\alpha}F\left(T /|\delta
|^{\nu z}\right),
\end{equation}
where $\delta=(H-H_{c2})$ as defined before. The Gaussian nature of
the exponents in $3d$ is a consequence that the effective dimension
$d_{eff}=d+z=5$, which is larger than the upper critical
dimension\cite{livroM}, $d_c=4$.

\emph{Critical line}---The critical line that separates the
polarized PARA state from the AF phase with nonzero staggered
magnetization can be written on the neighborhood of the QCP as,
$T_{N}(H)\propto |H_{c2}-H|^{\psi}$. Theories for a $3d$ Bose
gas\cite{Giamarchi,Nohadani} and mean-field treatment\cite{Nikuni}
give a universal value, $\psi=2/3$. Scaling theory shows that
although  $d_{eff}>d$ in three dimensions, the magnon-magnon
interaction is dangerously irrelevant and must be
considered\cite{Mucio3,Millis}. Then we expect that the quartic
corrections to the mean-field result Eq. (\ref{mf}) will be now
important. Within the H-P representation for the spin operators, the
mean-field Hamiltonian including the dynamical spin wave
interactions is given by,
\begin{equation}\label{mf1}
\mathcal{H}_{mf}^{'}=\sum_{k}\omega_{k}b_{k}^{\dagger}b_{k}+
\sum_{k,k'}\left(\frac{JZ}{2}+D\right)b_{k}^{\dagger}b_{k}b_{k'}^{\dagger}b_{k'}+E_{g},
\end{equation}
where $(\frac{JZ}{2}+D)$ works as an effective repulsion between the
magnons. The dispersion $\omega_k$ is given by Eq. (\ref{w}) and
$E_g$ by Eq. (\ref{eg}). We decouple the spin-wave interaction as,
$b_{k}^{\dagger}b_{k}b_{k'}^{\dagger}b_{k'} \approx
b_{k}^{\dagger}b_{k}<b_{k}^{\dagger}b_{k}>$.  Thus, trivially we
obtain the internal energy as,
\begin{equation}\label{U'}
U'=\sum_{q}\omega_{q}'<b_{q}^{\dagger}b_{q}>+E_{g},
\end{equation}
where we have already considered the proximity to the QCP
($H=H_{c2}$, $T=0$). The spectrum of excitations taking into account
magnon-magnon interactions is,
\begin{equation}\label{wh1}
\omega_{q}' =
(H-H_{c2})+\left(\frac{JZ}{2}+D\right)<b_{k}^{\dagger}b_{k}>+\mathfrak{D}q^{2}.
\end{equation}
We set up an equation for the critical line within the mean-field
approximation where the effect of magnon-magnon fluctuations are
included  in a self-consistent manner. The critical temperature
$T_N(H)$ is determined by the condition $\delta(H,T_N)=0$ where,
\begin{equation*}
\delta(H,T_N) = (H-H_{c2})           +
\end{equation*}
\begin{equation}
\left(\!\frac{JZ}{2}\!+\!D\!\right)\frac{S_{d}}{4\pi^{2}}\left(\frac{k_{B}T_{N}}{\mathfrak{D}}\right)^{d/2}\!\!
\int_{0}^{\infty}\!\!\!dxx^{\frac{d}{2}-1}\!\left(\coth\frac{x}{2}\!-\!1\right).
\end{equation}
For $2d$ the integral above diverges as expected from general
arguments\cite{Mermin}. For $3d$ we get, $
k_{B}T_{N}=(\zeta(3/2)(JZ/2+D))^{-2/3}
\pi\mathfrak{D}(H_{c2}-H)^{2/3}$. Notice that the effective
magnon-magnon coupling strength $(JZ/2+D)$ determines the transition
temperature, despite the Gaussian exponents, as expected from the
\emph{dangerously} irrelevant nature of the magnon-magnon
interactions. If we write the equation for the critical line,
$\delta(H,T)=0$, in the form, $H_{c2}(T)=H_{c2}(0)-v_0T^{1/\psi}$,
with $v_0$ related to the spin-wave interaction, we identify the
{\em shift exponent}, $\psi=z/(d+z-2)=2/3$, in agreement with the
renormalization group (RG) result\cite{Millis}. The temperature
dependence of $\delta$ arising from the spin-wave interactions can
modify the temperature dependence of $\Delta M$, $C_V$ and $\chi$ at
$H=H_{c2}$. In the limit $T \rightarrow 0$ we can easily see that
the purely Gaussian results for  $\Delta M$ and $C_V$ calculated
above are dominant. However, for the longitudinal susceptibility the
spin-wave interactions modify the purely Gaussian result. In this
case, it is straightforward to show that for, $H=H_{c2}$,
$T\rightarrow 0$, the dominant is $\chi \propto T^{1/4}$, instead of
$\chi \propto T^{1/2}$ calculated before.

\begin{figure}[th]
\centering
\includegraphics{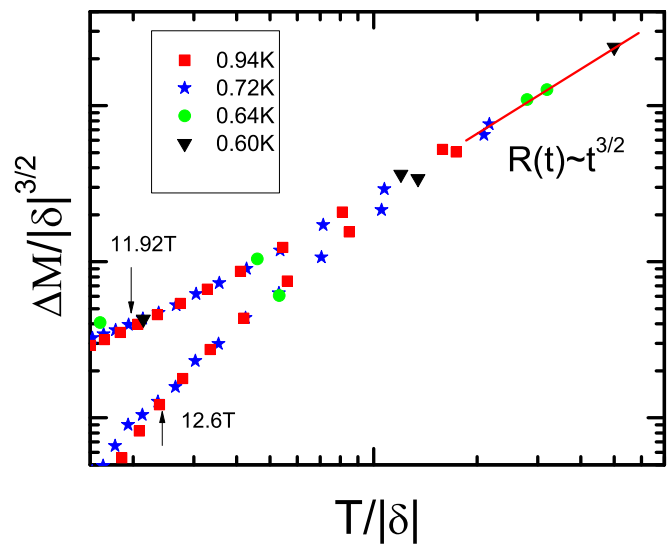}
\caption{(Color online) Scaling plot in logarithmic scales of the
magnetization data for the compound DTN obtained for fields up to
17T and temperatures $T=0.60, 0.64, 0.72$ and $0.94K$. The line
shows the asymptotic behavior of the scaling function $R(t) \sim
t^{3/2}$ in Eq. (\ref{scaling}), for $H \rightarrow H_{c2}$. The
arrows indicate the region of validity of the scaling. }\label{fig1}
\end{figure}
\emph{Scaling analysis of the Magnetization}---We start from the
free energy density, which close to the zero temperature quantum
phase transition has the scaling form given by Eq. (\ref{free}). The
zero temperature critical exponents $\alpha$, $\nu$ and the
dynamic exponent $z$ are related to the dimensionality of the
system $d$ by the quantum hyperscaling relation,
$2-\alpha=\nu(d+z)$\cite{livroM}. In general for $d_{eff}=d+z>4$,
i.e., above the upper critical dimension $d_{c}=4$, the exponents
associated with the QCP at $\delta=0$ take Gaussian values, and in
particular the correlation length exponent, $\nu=1/2$. That this
is the case in the present theory can be immediately verified
writing the thermodynamic functions in a scaling form and
identifying the relevant exponents. Furthermore Eq. (\ref{wh})
yields the dynamic exponent $z=2$. Using the relation $\Delta M
\propto
\partial f/\partial H$ we get,
\begin{equation}
\label{scaling} \Delta M \propto |\delta |^{1-\alpha}R\left(T/
|\delta |^{\nu z}\right),
\end{equation}
where $\Delta M=M_{sat}(T,H_{sat})-M(T,H)$ and $M_{sat}$ is measured
at the highest fields, $H_{sat}\gtrsim15$T. Using the hyperscaling
relation for $3d$ we obtain $1-\alpha=\nu(3+z)-1=(1/2)(3+2)-1=3/2$.

Figure \ref{fig1} shows the scaling plot of the magnetization for
the compound DTN in fields up to $17$T and for several temperatures.
The magnetization data was obtained using a vibrating sample
magnetometer adapted to be used in a $^3$He cryostat. The external
magnetic field, produced by a superconducting coil, was aligned with
the tetragonal axis of the sample, necessary condition to induce
BEC. As shown in the figure the experimental data collapses in a
good scaling plot when using the critical exponents appropriate for
three dimensions. It can also be seen in Fig. \ref{fig1} that for $H
\rightarrow H_{c2}$, the scaling function $R(t\rightarrow \infty)
\sim t^{3/2}$, such that, in this limit $\Delta M_{3d} \propto
T^{3/2}$ in agreement with the theory. The lower (upper) branch in
Fig. \ref{fig1} corresponds to data for higher (lower) fields than
the critical magnetic field $H_{c2}(0)=$12.3T. This critical field
obtained from a criterion of best data collapse is in very good
agreement with that found by Paduan-Filho et al.\cite{Paduan2} using
numerical differentiation of the magnetization data. We point out
that a good scaling of the data is observed for magnetic fields
sufficiently close to the critical field, i.e., for
11.92T$<H_{c2}<$12.6T.

\emph{Summary}---In spite of the large literature on the subject of
BEC of magnons, the phase diagram and the thermodynamic properties
around the upper critical magnetic field $H_{c2}$, have not yet been
completely examined. Since, in general, high magnetic fields are
necessary to reach $H_{c2}$, experimental results and consequently
theoretical work are much more scarce in this region of the phase
diagram. As we pointed out, the organic compound
$NiCl_2$-$4SC(NH_2)_2$ (DTN) is ideal for this kind of studies since
detailed magnetization curves can be obtained for very low
temperatures close to $H_{c2}$. With this motivation we introduced a
BOMFT approximation to study theoretically the upper field
transition. We have obtained the dominant low temperature behavior
of the magnetization ($\Delta M \propto T^{3/2}$), specific heat
($C_V \propto T^{3/2}$) and susceptibility ($\chi \propto T^{1/4}$)
at the quantum critical trajectory and determined the shift exponent
of the Neel line. We pointed out that, although the magnon-magnon
interactions are irrelevant in the RG sense close to the QCP, they
should be taken into account and determine the temperature
dependence of the critical line and that of the susceptibility along
the quantum critical trajectory. Our mean-field approach is
justified since the effective dimension for the transition at the
QCP ($H=H_{c2}$, $T=0$) is above the upper critical dimension.
Finally using the theoretical prediction we obtained for the scaling
form of the field and temperature dependent magnetization close to
$H=H_{c2}$, $T=0$, we performed a scaling analysis of our
magnetization data for DTN. The very good agreement between the
theoretical and experimental results provides unequivocal evidence
that the transition at  $H_{c2}$ is a BEC of magnons.

D. Reyes would like to thank Dr. Han-Ting Wang and Dr.  Stefan
Wessel for many illuminating discussions. Support from the Brazilian
agencies CNPq and FAPERJ is gratefully acknowledged.

\end{document}